\documentclass[sigconf]{acmart}

\usepackage{subfigure}
\usepackage{CJKutf8}
\usepackage{extarrows}
\usepackage{url}
\usepackage{booktabs}
\usepackage{bm}
\usepackage{enumitem}
\usepackage{multirow}

\AtBeginDocument{%
  }

\copyrightyear{2025}
\acmYear{2025}
\setcopyright{cc}
\setcctype{by}

\acmConference[SIGIR '25]{Proceedings of the 48th International ACM SIGIR Conference on Research and Development in Information Retrieval}{July 13--18, 2025}{Padua, Italy}
\acmBooktitle{Proceedings of the 48th International ACM SIGIR Conference on Research and Development in Information Retrieval (SIGIR '25), July 13--18, 2025, Padua, Italy}\acmDOI{10.1145/3726302.3730037}
\acmISBN{979-8-4007-1592-1/2025/07}

\begin{document}

\title{Mitigating Modality Bias in Multi-modal Entity Alignment from a Causal Perspective}

\author{Taoyu Su}
\orcid{0009-0003-1674-7635}
\affiliation{%
  \institution{Institute of Information Engineering, Chinese Academy of Sciences}
  \institution{School of Cyber Security, University of Chinese Academy of Sciences}
  \city{Beijing}
  \country{China}}
\email{sutaoyu@iie.ac.cn}

\author{Jiawei Sheng}\authornote{Corresponding author.}
\orcid{0000-0002-4865-982X}
\affiliation{%
  \institution{Institute of Information Engineering, Chinese Academy of Sciences}
  \city{Beijing}
  \country{China}}
\email{shengjiawei@iie.ac.cn}

\author{Duohe Ma}
\orcid{0000-0002-2160-8344}
\affiliation{%
  \institution{Institute of Information Engineering, Chinese Academy of Sciences}
  \city{Beijing}
  \country{China}}
\email{maduohe@iie.ac.cn}

\author{Xiaodong Li}
\orcid{0009-0008-7374-5413}
\affiliation{%
  \institution{Institute of Information Engineering, Chinese Academy of Sciences}
  \institution{School of Cyber Security, University of Chinese Academy of Sciences}
  \city{Beijing}
  \country{China}}
\email{lixiaodong@iie.ac.cn}

\author{Juwei Yue}
\orcid{0000-0002-6899-5724}
\affiliation{%
  \institution{Institute of Information Engineering, Chinese Academy of Sciences}
  \institution{School of Cyber Security, University of Chinese Academy of Sciences}
  \city{Beijing}
  \country{China}}
\email{yuejuwei@iie.ac.cn}

\author{Mengxiao Song}
\orcid{0009-0004-7763-0745}
\affiliation{%
  \institution{Institute of Information Engineering, Chinese Academy of Sciences}
  \institution{School of Cyber Security, University of Chinese Academy of Sciences}
  \city{Beijing}
  \country{China}}
\email{songmengxiao@iie.ac.cn}

\author{Yingkai Tang}
\orcid{0009-0001-3346-5512}
\affiliation{%
  \institution{Institute of Information Engineering, Chinese Academy of Sciences}
  \institution{School of Cyber Security, University of Chinese Academy of Sciences}
  \city{Beijing}
  \country{China}}
\email{tangyingkai@iie.ac.cn}

\author{Tingwen Liu}
\orcid{0000-0002-0750-6923}
\affiliation{%
  \institution{Institute of Information Engineering, Chinese Academy of Sciences}
  \institution{School of Cyber Security, University of Chinese Academy of Sciences}
  \city{Beijing}
  \country{China}}
\email{liutingwen@iie.ac.cn}

\renewcommand{\shortauthors}{Taoyu Su et al.}

\begin{abstract}
Multi-Modal Entity Alignment (MMEA) aims to retrieve equivalent entities from different Multi-Modal Knowledge Graphs (MMKGs), a critical information retrieval task.
Existing studies have explored various fusion paradigms and consistency constraints to improve the alignment of equivalent entities, while overlooking that the visual modality may not always contribute positively. 
Empirically, entities with low-similarity images usually generate unsatisfactory performance, highlighting the limitation of overly relying on visual features.
We believe the model can be biased toward the visual modality, leading to a shortcut image-matching task. 
To address this, we propose a counterfactual debiasing framework for MMEA, termed CDMEA, which investigates visual modality bias from a causal perspective. 
Our approach aims to leverage both visual and graph modalities to enhance MMEA while suppressing the direct causal effect of the visual modality on model predictions. 
By estimating the Total Effect (TE) of both modalities and excluding the Natural Direct Effect (NDE) of the visual modality, we ensure that the model predicts based on the Total Indirect Effect (TIE), effectively utilizing both modalities and reducing visual modality bias. 
Extensive experiments on 9 benchmark datasets show that CDMEA outperforms 14 state-of-the-art methods, especially in low-similarity, high-noise, and low-resource data scenarios.

\end{abstract}


\maketitle

\section{Introduction}\label{Sec:intro}

\begin{figure}
    \centering
    \includegraphics[width=1.0\linewidth]{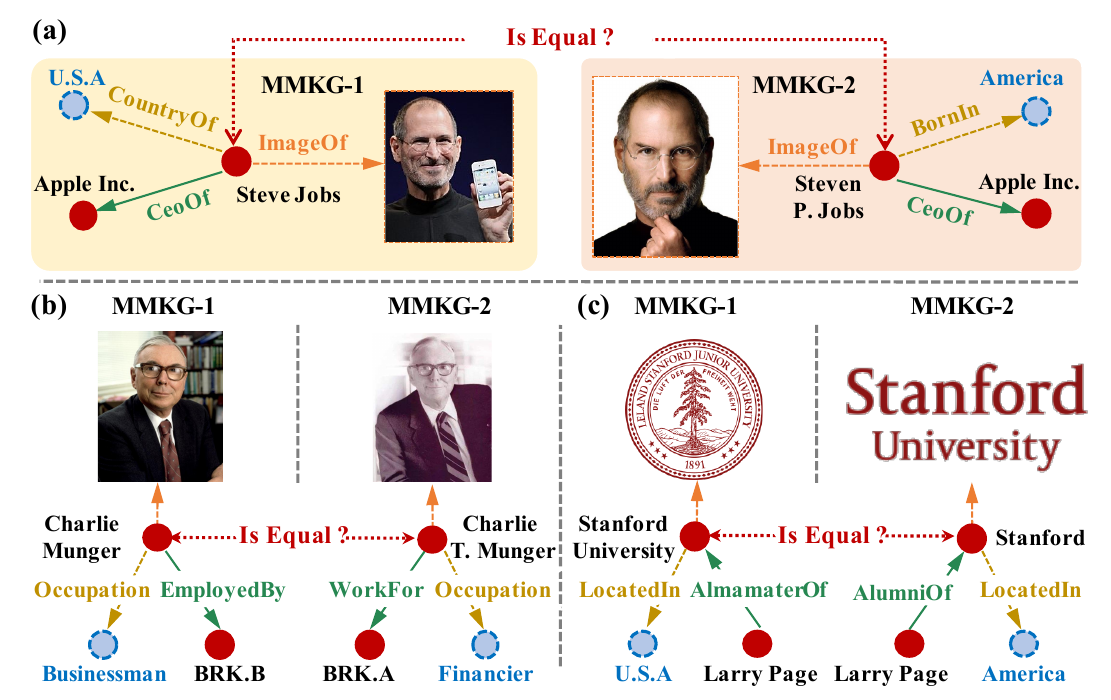}
    \caption{Examples of MMEA between two MMKGs: (a) entity equivalence identification, (b) MMEA with supportive visual cues, and (c) MMEA with negative visual cues.}
    \label{fig:intro}
\end{figure}

Identifying equivalent entities across Knowledge Graphs (KGs) is an intrinsic Information Retrieval (IR) problem~\cite{SEA_sigir,EasyEA_sigir,ER_sigir}, known as Entity Alignment (EA)~\cite{POE,MMKG_Survey}.
With the rise of Multi-Modal Knowledge Graphs (MMKGs), which enrich relational data with multi-modal information, Multi-Modal Entity Alignment (MMEA) has garnered significant attention.
As illustrated in Figure~\ref{fig:intro} (a), given an entity \texttt{Steve\_Jobs} in MMKG-1, an MMEA model is expected to retrieve \texttt{Steven\_Paul\_Jobs} as the equivalent entity from all candidate entities in MMKG-2, with utilization of contextual multi-modal data.
This task is foundational for various knowledge-based IR applications, such as image retrieval~\cite{DBLP:conf/sigir/WenSCWNC24,MMKG_Img_Search,MKG4Paper}, recommendation systems~\cite{MKG4rRS,KG4Rec_sigir}, and knowledge representation~\cite{MKG4EM,MKG4QA,NativE,CMR}. 
 
The key challenge in MMEA is facilitating consistency between equivalent entities across MMKGs. 
Recent studies have primarily focused on two aspects. 
One involves the design of \textit{multi-modal fusion} paradigms, such as feature concatenation~\cite{POE,HMEA,MMEA} and weighted attention mechanisms~\cite{EVA,MCLEA,UMAEA,PSNEA,Meaformer,XGEA,LoginMEA}, to extract multi-modal features for MMEA.
The other focuses on developing \textit{consistency constraints} between equivalent entities across MMKGs, using contrastive losses~\cite{MCLEA,ACK-MMEA,PMF,SimDiff} or other regularizers~\cite{PCMEA,IBMEA,GEEA,ASGEA}.
These methods improve MMEA but struggle with low-quality data~\cite{IBMEA}.
Incorporating images often enhances alignment, as shown in Fig.~\ref{fig:intro} (b), where the images of \texttt{Charlie\_Munger} share supportive features that aid in alignment.
However, when images of equivalent entities differ significantly, as in Fig.\ref{fig:intro} (c) with \texttt{Stanford\_University}, their low similarity can hinder consistency and impede MMEA.

\begin{figure}[t]
    \centering
    \includegraphics[width=1\linewidth]{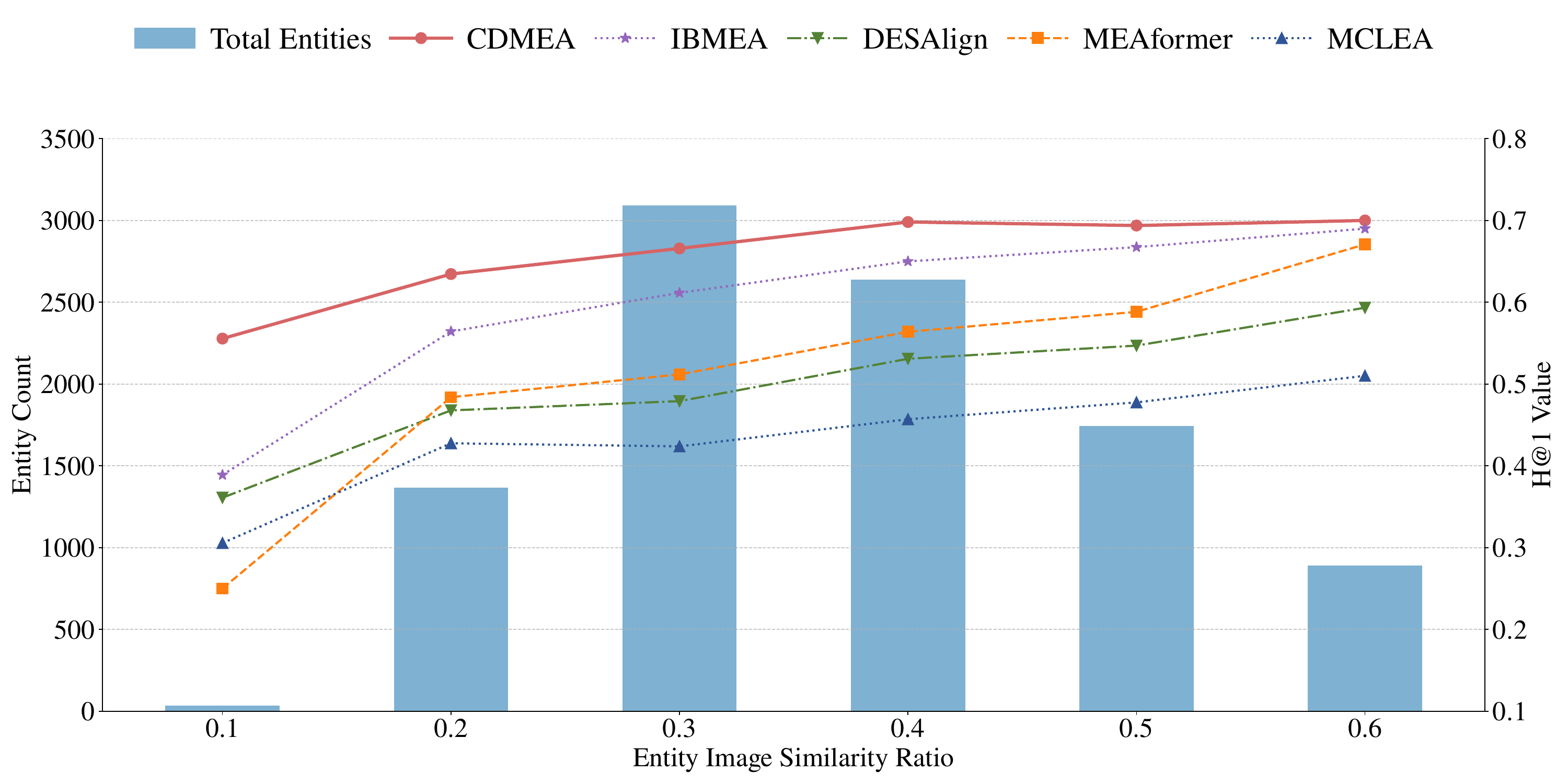}
    \caption{Experimental results on equivalent entity pairs with low-similarity images from the FB-DB15K dataset.}
    \label{fig:intro_low_img}
\end{figure}

To further investigate the impact of low-similarity images, we conduct experiments with advanced methods, shown in Fig.~\ref{fig:intro_low_img}.
Our statistics show that 86.3\% of equivalent entity pairs have image similarity below 0.5, and 43.7\% below 0.3, using the cosine similarity of image features from VGG-16~\cite{VGG16}, on the FB-DB15K test data~\cite{POE}.
The figure also shows that existing advanced methods perform poorly on low-similarity image data.
We believe that existing methods may overly rely on the automatically learned modality fusion modules, which can cause a bias toward the visual modality, leading to shortcut learning~\cite{shortcut_learing,IBMEA} that simplifies or degrades the MMEA task into an image-matching problem.

This motivates us to explore visual \textit{modality bias} in MMEA from a causal perspective~\cite{causal_learning}.
Specifically, we consider both the visual and graph modalities, where the graph modality includes entities' structures, relations, and attributes for simplicity.
As shown in Fig.~\ref{fig:intro} (c), the visual modality alone may mislead MMEA, while the graph modality provides useful cues.
The Total Effect (TE) of both modalities reflects overall predictive information, while the Natural Direct Effect (NDE) of the visual modality introduces bias.
To explore this, we introduce two hypothetical worlds of MMEA: 
\begin{itemize}[leftmargin=*]
    \item \textit{\textbf{Factual World of MMEA}: Prediction using both visual and graph modalities, representing the real MMEA task.}
    \item \textit{\textbf{Counterfactual World of MMEA}: Prediction using only the visual modality, simplifying MMEA to an image-matching task.}
\end{itemize}
The counterfactual world estimates the bias of the visual modality (i.e., NDE). 
We aim to exclude the NDE from the TE to perform counterfactual inference\cite{NDE}. 
In this way, \textit{we preserve positive predictive information from both modalities while mitigating the bias from the visual modality}, leading to a debiased MMEA prediction.

To achieve the above idea, we propose a general \underline{C}asual \underline{D}ebiasing framework for \underline{M}ulti-modal \underline{E}ntity \underline{A}lignment, termed as CDMEA\footnote{Our code is available at {\small\url{https://github.com/sutaoyu/CDMEA}}.}. 
In general, we estimate the Total Effect (TE) and Natural Direct Effect (NDE) from the factual and counterfactual worlds.
First, to derive TE, we extract visual, graph, and fused modality features using modality encoders and calculate the prediction score for each feature separately. 
By integrating the three predictions, we obtain the final prediction to estimate TE in the factual world.
Second, to estimate the NDE of the visual modality on the prediction, we mask the graph and the fused modality information to construct a counterfactual world. 
We then make a prediction using only the visual modality, which gives the NDE reflecting the bias toward the visual modality.
To obtain the final debiased prediction, we can subtract NDE from TE. 
Our major contributions are as follows:
\begin{itemize}[leftmargin=*]
    \item \textbf{Debiasing Principle}: We introduce a novel perspective of causal debiasing and propose a causal graph for MMEA.
    To the best of our knowledge, we are the first to comprehensively investigate the causal effects of both visual and graph modalities in MMEA. 
    \item \textbf{General Framework and Implementation}: We propose CDMEA, a causal debiasing framework for MMEA, and implement it using advanced techniques, including visual, graph, and fusion modality encoders, along with a prediction integration strategy.
    \item \textbf{Extensive Experiments}: Our experiments demonstrate that CDMEA outperforms 14 state-of-the-art methods on 9 benchmarks, showing promising and robust results in the low-similarity, high-noise, and low-resource data scenarios.
\end{itemize}

\section{Preliminaries}

\subsection{Task Formulation}

A \textbf{Multi-Modal Knowledge Graph (MMKG)} is defined as $\mathcal{K}=(\mathcal{E},\mathcal{R}, \mathcal{A}, \mathcal{V})$, where $\mathcal{E}$, $\mathcal{R}$, $\mathcal{A}$, and $\mathcal{V}$ represent the sets of entities, relations, attributes, and visual images, respectively. 
The relational triples, $\mathcal{T} \subseteq \mathcal{E} \times \mathcal{R} \times \mathcal{E}$, describe the relationships between entities. 
We refer to the contextual information of $\mathcal{T}$ and $\mathcal{A}$ as the \textit{graph modality} $G$, and the visual information $V$ as the \textit{visual modality}.

\textbf{Multi-Modal Entity Alignment (MMEA)} aims to retrieve equivalent entities across two MMKGs representing the same real-world concept.
Formally, given two MMKGs $\mathcal{K}^{(1)}$ and $\mathcal{K}^{(2)}$ with their relational triples and visual images of the entities, MMEA identifies equivalent entity pairs $\{\langle e^{(1)}, e^{(2)} \rangle | e^{(1)} \in \mathcal{K}^{(1)}, e^{(2)} \in \mathcal{K}^{(2)}, e^{(1)} \equiv e^{(2)}\}$, where $e^{(1)}$ and $e^{(2)}$ are entities from each MMKG. 
The model is trained on pre-aligned entity pairs $\mathcal{S}$, and during inference, it ranks candidate entities in one MMKG to predict the equivalent entity in the other. 
This work focuses on improving consistency in multi-modal entity alignment and mitigating visual modality bias to enhance MMEA performance.

\subsection{Causal Learning}

Causal learning~\cite{causal_learning} estimates causal effects on model predictions and is widely used in fields like information retrieval~\cite{Causal_retrival_sigir,Causal_Legal_Case_sigir,SCM4SR_sigir}.
A random variable represents an uncertain quantity (e.g., input feature or outcome), denoted by a capital letter (e.g., $X$), with its observed value denoted by a lowercase letter (e.g., $x$).

\subsubsection{Causal Graph}

\begin{figure}
    \centering
    \includegraphics[width=0.9\linewidth]{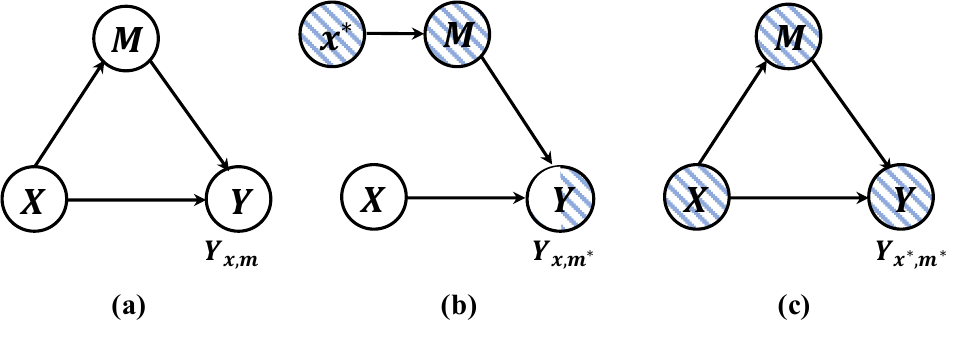}
    \caption{Example of the causal graph. (a) Factual world. (b, c) Counterfactual world. 
    White nodes represent observed variables, while blue-striped nodes are counterfactual variables.
    }
    \Description{Example of the causal graph.}
    \label{fig:Causal_graph_example}
\end{figure}

A causal graph~\cite{causal_learning} represents the causal relationships between variables as a directed acyclic graph $\mathcal{G}_{causal} = {\mathcal{N}, \mathcal{E} }$, where $\mathcal{N}$ is the set of variables and $\mathcal{E}$ is the set of causal relationships.
Fig.~\ref{fig:Causal_graph_example} (a) shows an example with three variables: $X$, $Y$, and $M$.
The variable $X$ has both a direct effect on $Y$ ($X \rightarrow Y$) and an indirect effect via $M$ ($X \rightarrow M \rightarrow Y$), where $M$ is a mediator. 
When $X = x$ and $M = m$, the value of $Y$ is:
\begin{equation}
    Y_{x,m} = Y(X = x, M = m),
\end{equation}
which reflects both effects. 
In the factual world~\cite{causal_learning}, $m = M_x = M(X = x)$, preserving the impact of $X \rightarrow M$.

To analyze causal effects, we present a \textit{counterfactual world}~\cite{causal_learning}, where $X$ can take different values while $M$ and $Y$ vary accordingly. 
For example, as shown in Fig.~\ref{fig:Causal_graph_example} (b), we can set $X=x$ and $M=m^*$, yielding the counterfactual value $Y_{x,m^*}=Y(X=x, M=m^*)$. 
In this case, $M$ is set to a different value $m^*=M(x=x^*)$. 
Another example of a counterfactual world is depicted in Fig.~\ref{fig:Causal_graph_example} (c).

\subsubsection{Casual Effects}\label{sec:casual_effects}

The causal effect quantifies the difference between two potential outcomes under different treatments~\cite{causal_learning, CF-VQA}. 
Let $X=x$ denote the individual under treatment and $X=x^*$ under no-treatment condition~\footnote{For example, $X=x$ means taking a drug, and $X=x^*$ means not taking it.}.
The \textit{Total Effect (TE)}~\cite{causal_learning} measures the total impact of treatment with $x$ compared to $x^*$:
\begin{equation}\label{eq:TE}
\begin{aligned}
TE = Y_{x,m}-Y_{x^*,m^*},
\end{aligned}
\end{equation}
where $Y_{x^*,m^*} = Y(X = x^*, M = m^*=M(X=x^*))$ denotes the outcome in counterfactual world.
TE captures both direct and indirect effects.
The \textit{Natural Direct Effect (NDE)}~\cite{causal_learning} isolates the direct causal impact of $X$ on $Y$ by blocking the indirect path (Fig.~\ref{fig:Causal_graph_example} (c)):
\begin{equation}\label{eq:NDE}
NDE = Y_{x,m^*} - Y_{x^*,m^*}.
\end{equation}
The \textit{Total Indirect Effect (TIE)}~\cite{causal_learning} measures the indirect influence of $X$ on $Y$ through $M$, excluding the direct effect.
While directly computing TIE can be complex due to multiple variables, it is often approximated by subtracting NDE from TE:
\begin{equation}\label{eq:TIE}
TIE = TE - NDE = Y_{x,m} - Y_{x,m^*}.
\end{equation}
TIE excludes the direct effect, thus mitigating biases caused by shortcuts from input to output~\cite{CCDF, CF-VQA, CDN}.
This insight motivates us to debias the MMEA model's visual modality.

\begin{figure}
    \centering
    \includegraphics[width=0.9\linewidth]{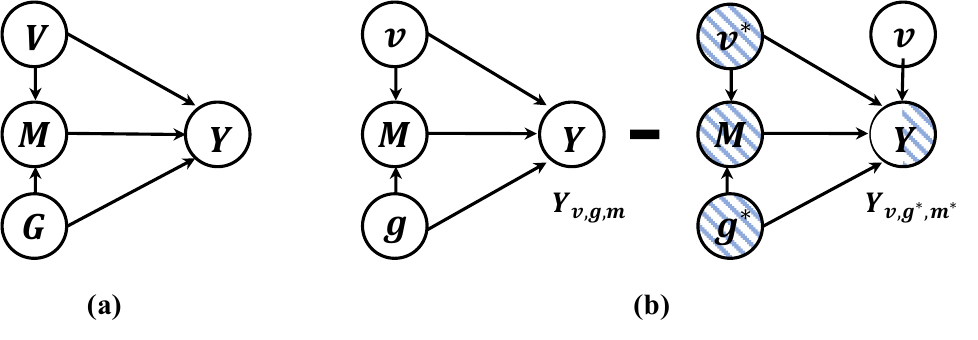}
    \caption{
    (a) Causal graph for MMEA. 
    (b) Counterfactual analysis illustrates the difference between factual and counterfactual inference outcomes for entities with observed values.
    }
    \label{fig:SCM}
\end{figure}

\begin{figure*}[!t]
    \centering
    \includegraphics[width=1.0\linewidth]{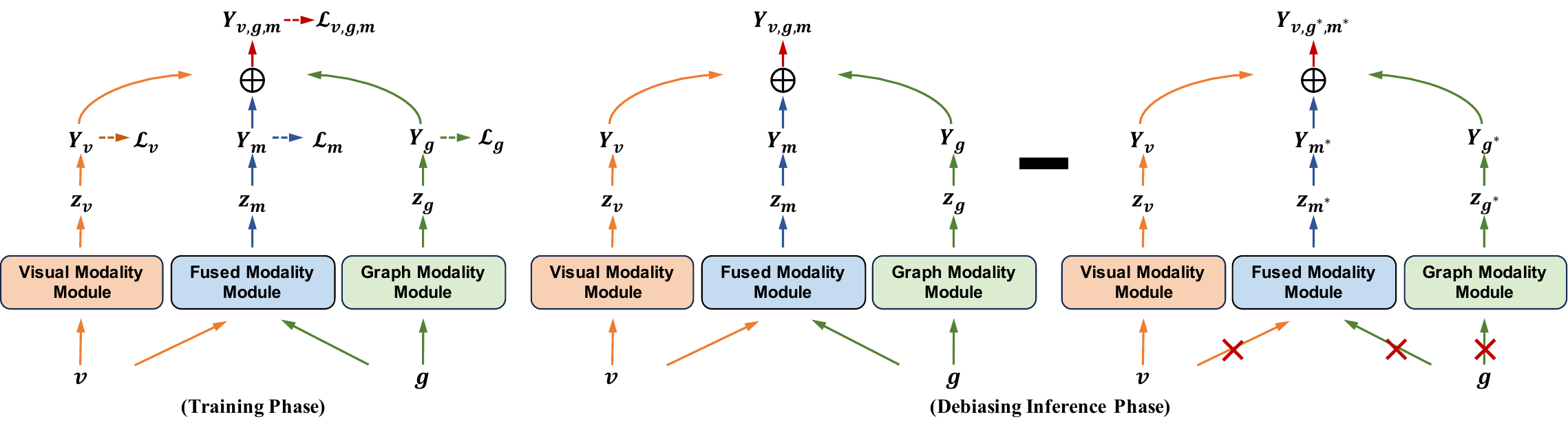}
    \caption{The framework of the proposed CDMEA in the training phase and debiasing inference phase.}
    \label{fig:CDMEA}
\end{figure*}

\section{Methodology}
This section constructs a causal graph to explore the modality bias in MMEA and its impact on performance.
We introduce the Counterfactual Debiasing framework (CDMEA) for MMEA, which mitigates modality bias to boost MMEA. 
A novel causal debiasing network for MMEA is then proposed to verify our approach.

\subsection{Causal Perspective of MMEA} \label{Sec:Causal_View_of_MMEA}

To investigate modality bias in MMEA, we construct a causal graph with four variables: visual modality $V$, graph modality $G$, fused modality $M$, and model prediction $Y$, as shown in Fig.~\ref{fig:SCM} (a). 
The nodes and links are defined as follows:

\begin{itemize}[leftmargin=*]
\item \textbf{Node $V$}. 
It denotes the information about the visual modality of the entity extracted from the entity's image.

\item \textbf{Node $G$}. 
It denotes the information about the graph modality of the entity derived from the MMKG topology. For simplicity, we summarize the connectivity, relation, and attribute information of the entity in the graph information.

\item \textbf{Node $M$}. 
It denotes the fused information of both visual and graph modality.
As a mediator, $M$ can be written as:
\begin{equation}\label{eq:node_F}
M_{v,g} = M(V=v, G=g).
\end{equation}

\item \textbf{Link $V\to M$ and $G\to M$}.
They refer to the direct causal effects of $V$ and $G$ on $M$, whereas the fused modality is built with the visual and the graph modality.

\item \textbf{Link $V\to Y$, $G\to Y$, and $M\to Y$}.
They refer to the direct causal effects of $V$, $G$, and $M$ on $Y$. 
In other words, it reflects the fact that the model prediction is decided by visual, graph, and fused-modality features.

\item \textbf{Node $Y$}.
It denotes the final model prediction, which combines the model predictions by three links $\{V, G, M\} \rightarrow Y$.
In the factual world, all input variables can be observed, which is:
\begin{equation}\label{eq:node_Y}
Y_{v,g,m} = Y(V=v, G=g, M=m),
\end{equation}
where $m=M_{v,g}$ denotes the fused modality information from both the entity's visual and graph modality.
\end{itemize}

\subsection{Counterfactual Debiasing Framework}
To mitigate visual modality bias, we propose CDMEA, a causal debiasing framework for MMEA. 
The goal is \textit{to leverage both visual and graph modalities while suppressing the direct causal effect of the visual modality on model predictions}.
This ensures that both modalities are fully utilized, preventing degradation in MMEA inference.

First, we estimate the total effect (TE) of MMEA by comparing the factual world and the counterfactual world:
\begin{equation}\label{eq:TE-new}
TE=Y_{v,g,m} - Y_{v^*,g^*,m^*},
\end{equation}
where $Y_{v,g,m}$ is the prediction in the factual world, and $Y_{v^*,g^*,m^*}$ in the counterfactual world.
Here, $m^*=M(V=v^*, G=g^*)$ is the counterfactual fused modality,
and  $v^*,g^*,m^*$ are dummy values, typically set to zero, resulting in a zero similarity score in MMEA.

Next, we derive the total indirect effect (TIE), which considers both visual and graph modality with interactions to generate a debiased prediction.
Estimating TIE directly is challenging, so we instead estimate the natural direct effect (NDE) of the visual modality and subtract it from TE to obtain TIE:
\begin{equation}\label{eq:NDE-new}
NDE= Y_{v,g^*,m^*}-Y_{v^*,g^*,m^*},
\end{equation}
where only the branch $V \to Y$ is maintained in the counterfactual world of $Y_{v,g^*,m^*}$.
Finally, we can immediately derive the TIE as:
\begin{equation}\label{eq:TIE_MMEA}
TIE=TE-NDE=Y_{v,g,m}-Y_{v,g^*,m^*}.
\end{equation}
The above procedure is illustrated in Fig.~\ref{fig:SCM} (b).
To refine TIE estimation, inspired by studies~\cite{CDN}, we introduce a factor $\beta$ to control the proportion of NDE excluded from TE:
\begin{equation}\label{eq:TIE_alpha}
TIE =Y_{v,g,m}-\beta \cdot Y_{v,g^*,m^*},
\end{equation}
where adjusting $\beta$ allows us to control the direct effect of the visual modality in the final prediction.

\subsection{Framework Implementation}

In this section, we propose the implementation of CDMEA.
The overall framework is shown in Fig.~\ref{fig:CDMEA}.

\subsubsection{Visual Modality Module}
This module aims to achieve the variable $V$ in the causal graph.
For the image $v$ of an entity, the visual modality representation $\boldsymbol{z}_v$ can be generated as follows:
\begin{equation}\label{Eq:Z_v}
\boldsymbol{z}_v = \mathcal{F}_{V}(v).
\end{equation}
To ensure fair comparison, we follow previous studies~\cite{EVA,MCLEA,Meaformer} and implement the visual encoder $\mathcal{F}_{V}$ by a pre-trained model (PVM) such as VGG-16~\cite{VGG16} or ResNet-152~\cite{Resnet152} to extract image embeddings.
The above module can be denoted as $\boldsymbol{z}_v = \mathcal{F}_{V}(z) := \mathrm{PVM}(v; \theta_{V})$ with the parameter $\theta_{V}$ shared between different MMKGs.

\subsubsection{Graph Modality Module}

This module aims to achieve the variable $G$ in the causal graph.
Given an entity with its contextual graph structure $g$ and its attributes $a$, the graph modality representation $\bm{z}_g$ is formulated as follows:
\begin{equation}\label{Eq:Z_g}
\boldsymbol{z}_g  = \mathcal{F}_{G}(g, a),
\end{equation}
where $\mathcal{F}_{G}$ is a graph encoder that generates structural entity features from relational neighbors.
Here, the graph structure $g$ is reflected by relational triples.
In addition, we treat attributes $a$ as the initial features of nodes (i.e., entities), which are represented by bag-of-attribute vectors, inspired by previous studies~\cite{DESAlign,Meaformer,IBMEA}.

Vanilla Graph Neural Networks (GNNs), such as GCN~\cite{GCN} or GAT~\cite{GAT}, can serve as the graph encoder but struggle with capturing diverse relations.
Inspired by studies~\cite{RREA,Dual-AMN}, we advise a \textit{relational reflection graph attention network (RRGAT)} to aggregate neighboring entities while preserving relational structure:
\begin{equation}
\begin{aligned}
\boldsymbol{h}_{e_{i}}^{l+1}&=\tanh(\sum_{e_{j} \in \mathcal{N}_{e_{i}}} \sum_{r_{k} \in \mathcal{R}_{i j}} \alpha_{i j k}^{l}\bm{W}_{r_k}\boldsymbol{h}_{e_{j}}^{l}),\\
\alpha_{i j k}^{l} &:=\frac{\exp (\boldsymbol{q}^{T}\boldsymbol{h}_{r_{k}})}{\sum_{{e_{j}}{'} \in \mathcal{N}_{e_{i}}} \sum_{{r_{k}}{'} \in \mathcal{R}_{i j{'}}} \exp (\boldsymbol{q}^{T}\boldsymbol{h}_{{r_{k}}{'}}))},\\
\bm{W}_{r_k} &:= I - 2\bm{h}_{r_k} \bm{h}_{r_k}^T, \label{Eq:RRGAT}
\end{aligned}
\end{equation}
where $\mathcal{N}_{e_{i}}$ is the set of neighbors of $e_i$,and $\mathcal{R}{i j}$ is the relation set between $e_i$ and its neighbors.
$\boldsymbol{h}_{e_{j}}^{l} \in \mathbb{R}^d$ is the embedding of $e_j$ at $l$-layer, initialized from attributes.
The relation embedding $\boldsymbol{h}{r_k}$ is normalized such that $||\boldsymbol{h}{r_k}||_2 = 1$.
There are two beneficial properties:
(1) \textit{Diverse Neighbor Importance}: The attention weight $\alpha_{i j k}^{l}$ adjusts the contribution of each neighbor $e_j\in\mathcal{N}_{e_{i}}$ based on its relational context, and $\boldsymbol{q}$ denotes a learnable vector.
(2) \textit{Relational Dimensional Isometry}: $\bm{W}_{r_k}$ preserves entity embedding's norm and relative distance, ensuring relational consistency after transformation~\cite{RREA}.
To achieve this, the weight $\bm{W}_{r_k}$ ought to be retained as an orthogonal matrix as in Eq.~(\ref{Eq:RRGAT}), which can be easily proved as:
\begin{equation}
\begin{aligned}
    \bm{W}_{r_k}^T \bm{W}_{r_k} &= \left(I - 2 \bm{h}_{r_k} \bm{h}_{r_k}^T\right)^T \left(I - 2 \bm{h}_{r_k} \bm{h}_{r_k}^T\right) \\
    &= I - 4 \bm{h}_{r_k} \bm{h}_{r_k}^T + 4 \bm{h}_{r_k} \bm{h}_{r_k}^T \bm{h}_{r_k} \bm{h}_{r_k}^T = I.
\end{aligned}
\end{equation}

To further capture multi-hop neighborhood information, we stack outputs of multiple layers in RRGAT.
Therefore, the graph modality representation $\boldsymbol{z}_{g}$ can be derived as: 
\begin{equation}
\begin{aligned}\label{Eq:Z_g_achieve}
\boldsymbol{z}_{g}=\left[\boldsymbol{h}_{e}^{0}\| \boldsymbol{h}_{e}^{1}\|\ldots\| \boldsymbol{h}_{e}^{l}\right],
\end{aligned}
\end{equation}
where $\|$ denotes concatenation.
We briefly denote the RRGAT module above by $\boldsymbol{z}_g  = \mathcal{F}_{G}(g,a) := \mathrm{RRGAT}(g, a; \theta_{G})$ with the learnable parameter $\theta_{G}$ shared between MMKGs.

\subsubsection{Fused Modality Module}
This module aims to achieve the variable $M$ in the causal graph.
To utilize both visual and graph modality information for MMEA, we propose a fused modality module.
Given the visual $v$ and graph $g$ information, the fused modality representation $\boldsymbol{z}_m$ of an entity is defined as:
\begin{equation}\label{Eq:Z_m}
\boldsymbol{z}_m  = \mathcal{F}_{M}(g, v),
\end{equation}
where $\mathcal{F}_{M}$ is a general modality fusion function.

To facilitate modality fusion, inspired by XGEA~\cite{XGEA}, we employ a GNN encoder to refine visual data through relational graph structures via message passing.
For convenience and beneficial properties, we also employ RRGAT to achieve this fusion, as:
\begin{equation}
\begin{aligned}\label{Eq:Z_m_achieve}
\boldsymbol{z}_m  = \mathcal{F}_{M}(g, v) := {\mathrm{RRGAT}}(g, v; \theta_{M}),
\end{aligned}
\end{equation}
where the images of entities are input as the initial features of nodes, and $\theta_{M}$ is the learnable parameter.
This approach not only naturally enables cross-modal interactions between visual and graph modalities, but also ensures relational consistency for entities across MMKGs through the shared RRGAT.

\subsubsection{Alignment Prediction Module}
To achieve MMEA, we design an alignment prediction module to compute semantic similarity scores between entities from different MMKGs.
Given entity $e^{(1)}\in \mathcal{K}^{(1)}$ and $e^{(2)}\in \mathcal{K}^{(2)}$, we calculate the similarity scores for the visual, graph, and fused modalities as follows:
\begin{equation}
\begin{aligned}\label{Eq:Y_single}
    &\begin{cases} 
    Y_v = \mathcal{F}_{sim} (\bm{z}_{v}^{(1)}, \bm{z}_{v}^{(2)}) & \text{if } V = v \\
    Y_{v^{*}} = 0 & \text{if } V = \varnothing 
    \end{cases},\\
    &\begin{cases} 
    Y_g = \mathcal{F}_{sim} (\bm{z}_{g}^{(1)}, \bm{z}_{g}^{(2)}) & \text{if } G = g \\
    Y_{g^{*}} = 0 & \text{if } G = \varnothing 
    \end{cases},\\
    &\begin{cases} 
    Y_m = \mathcal{F}_{sim} (\bm{z}_{m}^{(1)}, \bm{z}_{m}^{(2)}) &\!\! \text{if } M = m \\
    Y_{m^{*}} = 0 &\!\! \text{if } M = \varnothing 
    \end{cases},\\
\end{aligned}
\end{equation}
where $\bm{z}_{v}$, $\bm{z}_{g}$ and $\bm{z}_{m}$ are the modality-specific representations from 
Eq.~(\ref{Eq:Z_v}), Eq.~(\ref{Eq:Z_g_achieve}) and Eq.~(\ref{Eq:Z_m_achieve}), respectively.
These representations are encoded separately from $\mathcal{K}^{(1)}$ and $\mathcal{K}^{(2)}$ using shared encoders.
The similarity function $\mathcal{F}_{sim}$ is implemented as the dot product, which is simple yet effective in our experiments.
Besides, we set the $Y_{v^{*}}$, $Y_{g^{*}}$, and $Y_{m^{*}}$ as zero, since if the corresponding input modality is blocked, indicating no meaningful similarity.

\subsubsection{Prediction Fusion Strategy}

To obtain the final prediction, we combine the scores $Y_{v}$,  $Y_{g}$, and $Y_{m}$ into the fused score $Y_{v,g,m}$ as:
\begin{equation}\label{eq:Y_final}
Y_{v,g,m} =\mathcal{F}_{fusion} (Y_v, Y_g, Y_m),
\end{equation}
where $\mathcal{F}_{fusion}$ can be implemented by simple sum or average strategy.
To capture the varying contributions of different modality features for MMEA, we utilize an attentive weighted sum strategy: 
\begin{equation}
    Y_{v,g,m}  = \sum_{k \in {v,g,m}}  \frac{\exp(\phi_k)}{\sum_{k^{'} \in {v,g,m}} \exp(\phi_{k^{'}})} Y_k, 
\end{equation}
where $\phi_{k}\in \mathbb{R}$ are modality-specific learnable weight parameters that control each modality's contribution to the final score.
Unlike previous studies that apply modality weights to input features~\cite{EVA,PSNEA}, we apply them to prediction scores within the debiasing framework.
In addition, we can also derive $Y_{v^*,g^*,m^*}$ and $Y_{v,g^*,m^*}$ in this way, facilitating the estimation of causal effects.

\begin{table*}[!th]
\footnotesize
\centering
\caption{
Experimental results on 2 cross-KG datasets, where X\% represents the percentage of seed alignments used for training.
The best result is \textbf{bold-faced} and the runner-up is \underline{underlined}. 
$*$ indicates results reproduced using the official source code.
}
\label{fbdb_result}
\setlength\tabcolsep{4.8pt}
{
{
\begin{tabular}{@{}lccccccccc|ccccccccc@{}}
\toprule
\multirow{2.5}{*}{\bf Methods} & \multicolumn{3}{c}{\bf FB-DB15K (20\%)} & \multicolumn{3}{c}{\bf FB-DB15K (50\%)} &  \multicolumn{3}{c|}{\bf FB-DB15K (80\%)} & \multicolumn{3}{c}{\bf FB-YG15K (20\%)} & \multicolumn{3}{c}{\bf FB-YG15K ((50\%)} &  \multicolumn{3}{c}{\bf FB-YG15K (80\%)}   \\

\cmidrule(r){2-4}\cmidrule(r){5-7}\cmidrule(r){8-10} \cmidrule(r){11-13}\cmidrule(r){14-16}\cmidrule(r){17-19}
& H@1  & H@10   &  MRR  
& H@1  & H@10   &  MRR  
& H@1  & H@10   &  MRR     
& H@1  & H@10   &  MRR     
& H@1  & H@10   &  MRR     
& H@1  & H@10   &  MRR     
\\

\midrule
TransE~\cite{TransE}  
& .078  & .240   &  .134   
& .230 & .446   & .306  
& .426 & .659   & .507 
& .064 & .203   &  .112   
& .197 & .382   &  .262  
& .392 & .595   &  .463     \\

IPTransE~\cite{IPTransE}  
& .065 & .215   &  .094   
& .210 & .421   &  .283  
& .403 & .627   &  .469     
& .047  & .169   & .084  
& .201  &  .369   & .248  
& .401 & .602   & .458    \\

GCN-align~\cite{GCN-Align}  
& .053 &  .174   &  .087   
& .226 &  .435   &  .293  
& .414 &  .635   &  .472     
& .081  & .235   & .153   
& .235  & .424   & .294  
& .406  & .643   & .477     \\

KECG~\cite{KECG}  
& .128 &  .340   &  .200   
& .167 &  .416   &  .251  
& .235 &  .532   &  .336     
& .094   & .274   &  .154   
& .167  & .381   & .241  
& .241  & .501   & .329     \\

\midrule

POE~\cite{POE} 
& .126 &  .151   &  .170   
& .464 &  .658   &  .533  
& .666 &  .820   &  .721     
& .113 &  .229   & .154   
& .347 &  .536   & .414  
& .573 &  .746   & .635     \\

\citeauthor{MMEA}~\cite{MMEA} 
& .265 &  .541   &  .357   
& .417 &  .703   &  .512  
& .590 &  .869   &  .685     
& .234 &  .480   &  .317   
& .403 &  .645   &  .486  
& .598 &  .839   &  .682     \\

HMEA~\cite{HMEA}
& .127 &  .369   &  -   
& .262 &  .581   &  -  
& .417 &  .786   &  -     
& .105 &  .313   &  -   
& .265 &  .581   &  -  
& .433 &  .801   &  -     \\

EVA~\cite{EVA}  
& .134 &  .338   &  .201   
& .223 &  .471   &  .307  
& .370 &  .585   &  .444     
& .098 &  .276   &  .158   
& .240 &  .477   &  .321  
& .394 &  .613   &  .471     \\

MSNEA~\cite{MSNEA}
& .114 &  .296   &  .175   
& .288 &  .590   &  .388  
& .518 &  .779   &  .613     
& .103 &  .249   &  .153   
& .320 &  .589   &  .413  
& .531 &  .778   &  .620     \\

ACK-MMEA~\cite{ACK-MMEA}  
& .304 &  .549   &  .387   
& .560 &  .736   &  .624  
& .682 &  .874   &  .752     
& .289 &  .496   &  .360   
& .535 &  .699   &  .593
& .676 &  .864   &  .744     \\

XGEA*~\cite{XGEA}  
& .475 &  .739   &  .565   
& .681 &  .857   &  .746  
& .791 &  .919   &  .840     
& .431 &  .691   &  .521   
& .585 &  .801   &  .666
& .705 &  .873   &  .768     \\

UMAEA*~\cite{UMAEA}
& .560 &  .719   &  .617   
& .701 &  .801   &  .736
& .789 &  .866   &  .817     
& .486 &  .642   &  .540
& .600 &  .726   &  .644 
& .695 &  .798   &  .732     \\

MCLEA*~\cite{MCLEA}  
& .447 &  .715   &  .538   
& .626 &  .838   &  .700
& .752 &  .900   &  .807     
& .407 &  .660   &  .494
& .541 &  .762   &  .618 
& .652 &  .847   &  .721     \\

DESAlign*~\cite{DESAlign}  
& .541 &  .811   &  .643   
& .706 &  .877   &  .770
& .824 &  .931   &  .863     
& .452 &  .712   &  .547
& .605 &  .802   &  .677 
& .727 &  .876   &  .780     \\

MEAformer*~\cite{Meaformer}  
& .558 & .807   &  .647   
& .698 & .871   &  .762  
& .794 & .918   &  .841     
& .449 &  .672   &  .527   
& .593 &  .789   &  .664
& .713 &  .852   &  .766     \\

PCMEA*~\cite{PCMEA}  
& .596 &  .819   &  .675  
& .699 &  .879   &  .765
& .809 &  \underline{.936}   &  .856     
& .518 &  \underline{.747}   &   \underline{.597}
& .615 &  \underline{.822}   &  .689 
& .721 &  \underline{.898}   &  .785     \\

SimDiff~\cite{SimDiff}  
& .615 &  \underline{.820}   &  .678   
& .731 &  \underline{.880}   &  .786
& \underline{.829} &  .929   &  \underline{.865}     
& \underline{.530} &  .736   &  .595
& \underline{.659} &  .820   &  \underline{.716} 
& .743 &  .886   &  .791     \\

IBMEA~\cite{IBMEA}
& \underline{.631} & .813   & \underline{.697}   
& \underline{.742} & \underline{.880}   & \underline{.793} 
& .821 & .922   & .859  
& .521   &  .708   & .584  
& .655    &  .821   &  .714  
& \underline{.751}   &  .890   &  \underline{.800}     \\

\midrule
\textbf{CDMEA} 
& \textbf{.674} & \textbf{.861}   & \textbf{.741}  
& \textbf{.785} & \textbf{.881}   & \textbf{.821} 
& \textbf{.835} & \textbf{.947}   & \textbf{.878}     
& \textbf{.623}   &  \textbf{.811}   &  \textbf{.690}  
& \textbf{.718}    &  \textbf{.877}   &  \textbf{.775}  
& \textbf{.793}   &  \textbf{.927}   &  \textbf{.843}     \\

\textit{\textcolor{blue}{Improv. \%}}
& \textcolor{blue}{$\uparrow$4.3}	& \textcolor{blue}{$\uparrow$4.1}	& \textcolor{blue}{$\uparrow$4.4}
& \textcolor{blue}{$\uparrow$4.3}	& \textcolor{blue}{$\uparrow$0.1}	& \textcolor{blue}{$\uparrow$2.8}
& \textcolor{blue}{$\uparrow$0.6}	& \textcolor{blue}{$\uparrow$1.1}	& \textcolor{blue}{$\uparrow$1.3}
& \textcolor{blue}{$\uparrow$9.3}	& \textcolor{blue}{$\uparrow$6.4}	& \textcolor{blue}{$\uparrow$9.3}
& \textcolor{blue}{$\uparrow$5.9}	& \textcolor{blue}{$\uparrow$5.5}	& \textcolor{blue}{$\uparrow$5.9}
& \textcolor{blue}{$\uparrow$4.2}	& \textcolor{blue}{$\uparrow$2.9}	& \textcolor{blue}{$\uparrow$4.3}
\\

\bottomrule

\end{tabular}
}}
\end{table*}

\subsection{Training and Inference}
After implementing the overall framework, we introduce the CDMEA training and inference phase.

\subsubsection{Model Training Phase}
To optimize the overall CDMEA framework, we impose training losses on all prediction scores $Y_v$, $Y_g$, $Y_m$ and $Y_{v,g,m}$, enabling causal effects of different branches of the causal graph (in Section~\ref{Sec:Causal_View_of_MMEA}).
The overall training loss is formulated as:
\begin{equation}\label{eq:Loss_all}
\begin{aligned}
\mathcal{L} &= \mathcal{L}_{v,g,m} + \mathcal{L}_{v} + \mathcal{L}_{g} + \mathcal{L}_{m},
\end{aligned}
\end{equation}
where $\mathcal{L}_{v,g,m}$, $\mathcal{L}_v$, $\mathcal{L}_g$, and $\mathcal{L}_m$ are losses over the branch of $Y_{v,g,m}$ $Y_v$, $Y_g$, and $Y_m$, respectively.
We implement all losses using the InfoNCE loss~\cite{InfoNCE} with the formulation as:
\begin{equation}\label{eq:Loss_k}
\begin{aligned}
\mathcal{L}_{k}= \sum_{(e_i,e_j) \in \mathcal{S}} -\log \frac{\exp (Y_{k}(e_i,e_j) / \tau)}{ \sum_{(e_i,e_{j^{'}}) \notin \mathcal{S}} \exp (Y_{k}(e_i,e_{j^{'}})) / \tau)},
\end{aligned}
\end{equation}
where $\mathcal{L}_{k}\in \{\mathcal{L}_{v,g,m}, \mathcal{L}_{v}, \mathcal{L}_{g}, \mathcal{L}_{m}\}$, $\tau\in \mathbb{R}$ is the temperature coefficient, and $\mathcal{S}$ is the pre-aligned equivalent entity pairs (i.e., seed alignments).
In this way, each modality-specific branch can make MMEA prediction, which can be used to debias inference.

\subsubsection{Debiasing Inference Phase}
In the inference phase, we perform a counterfactual debiasing inference using TIE as the criterion, defined in Eq.~(\ref{eq:TIE_alpha}). 
The TIE is formulated as:
\begin{equation}\label{eq:debias_inference}
\begin{aligned}
TIE &= Y_{v,g,m}-\beta \cdot Y_{v,g^*,m^*} \\
    &= \mathcal{F}_{fusion} (Y_v, Y_g, Y_m) - \beta \cdot \mathcal{F}_{fusion} (Y_v, Y_{g^*}, Y_{m^*}).
\end{aligned}
\end{equation}
Here, we introduce the factor $\beta\in \mathbb{R}$ to control the proportion of debiasing the visual modality. 
Given a target entity in one MMKG and all candidate entities in another MMKG, we select the entity with the highest TIE value of the candidate entities as the final prediction of the aligned entity.
In this way, by blocking the direct effect $V \rightarrow Y$, the model takes full modality utilization by path $V \rightarrow M \rightarrow Y$, thus achieving debiased MMEA.

\section{Experiments}
\subsection{Experimental Settings}
\subsubsection{Datasets and Evaluation Metrics}
We use two types of MMEA datasets in our experiments:  
(1) \textbf{Cross-KG Datasets}~\cite{POE}: We select the FB15K-DB15K (FB-DB15K for short) and FB-YAGO15K (FB-YG15K for short) datasets, which contain 128,486 and 11,199 labeled pre-aligned entity pairs, respectively.  
(2) \textbf{Bilingual Datasets}: DBP15K~\cite{JAPE,EVA} is a commonly used benchmark for bilingual entity alignment, consisting of three datasets derived from multilingual versions of DBpedia: $\mathbf{DBP15K}_{\rm ZH\mbox{-}EN}$, $\mathbf{DBP15K}_{\rm JA\mbox{-}EN}$, and $\mathbf{DBP15K}_{\rm FR\mbox{-}EN}$. 
Each bilingual dataset contains approximately 400K triples and 15K pre-aligned entity pairs.  
We utilize 20\%, 50\%, and 80\% of true entity pairs as alignment seeds for training on cross-KG datasets, and 30\% for bilingual datasets.
For evaluation, we use standard metrics including Hits@1 (H@1), Hits@10 (H@10), and Mean Reciprocal Rank (MRR).
Hits@N represents the proportion of correct entities ranked within the top N positions, while MRR calculates the average reciprocal rank of correct entities. 
For both metrics, higher values indicate better alignment performance.

\subsubsection{Baselines}
To evaluate the effectiveness of CDMEA, we compare it with several EA methods, categorized into two groups:
(1) \textbf{Traditional EA Methods}:  
We select 6 prominent entity alignment methods that rely solely on graph structures, excluding multi-modal information: TransE~\cite{TransE}, IPTransE~\cite{IPTransE}, GCN-align~\cite{GCN-Align}, KECG~\cite{KECG}, BootEA~\cite{BootEA}, and NAEA~\cite{NAEA}.
(2) \textbf{MMEA Methods}:  
We collect 14 state-of-the-art MMEA methods, which integrate entity images to enhance entity representations. These methods include POE~\cite{POE}, \citeauthor{MMEA}~\cite{MMEA}, HMEA~\cite{HMEA}, EVA~\cite{EVA}, MSNEA~\cite{MSNEA}, ACK-MMEA~\cite{ACK-MMEA}, XGEA~\cite{XGEA}, UMAEA~\cite{UMAEA}, MCLEA~\cite{MCLEA}, DESAlign~\cite{DESAlign}, MEAformer~\cite{Meaformer}, PCMEA~\cite{PCMEA}, SimDiff~\cite{SimDiff}, and IBMEA~\cite{IBMEA}. For further details, refer to Sec.\ref{sec:related_works}.  
We select MCLEA, DESAlign, MEAformer, and IBMEA as competitive methods for comparison.

\subsubsection{Implementation Details}
In our experiments, the graph encoder has a hidden layer and an output size of 300 for two layers. 
The visual feature dimension is 4096, with an embedding size of 100.
Training is conducted over 200 epochs with a batch size of 1,000, using the AdamW optimizer~\cite{AdamW} and a learning rate of 5e-4.
The hyperparameter $\beta$ in Eq.(\ref{eq:debias_inference}) is tuned from 0.0 to 0.9, yielding optimal results at 0.2. 
For further details about $\beta$, refer to Sec.~\ref{sec:alpha}.
For entities without images, we assign random visual modality vectors~\cite{MCLEA, UMAEA}, as done in previous works~\cite{Meaformer, IBMEA}. 
To address limited training data and ensure a fair comparison, we adopt an iterative training strategy and exclude entity names, following prior works~\cite{EVA, MCLEA, Meaformer}. 
All experiments are run on a 64-bit machine with two NVIDIA V100 GPUs.
The best hyperparameters are selected via grid search based on the H@1 metric.

\begin{table}[!t]
    \footnotesize
    \centering
    \caption{Experimental results on 3 bilingual datasets.
    }
    \label{fbdbp_result}
    \setlength\tabcolsep{2.5pt}
    {
    \begin{tabular}{@{}lccccccccc@{}}
    \toprule
    \multirow{2.5}{*}{\bf Methods} & \multicolumn{3}{c}{$\mathbf{DBP15K}_{\rm ZH\mbox{-}EN}$} & \multicolumn{3}{c}{$\mathbf{DBP15K}_{\rm JA\mbox{-}EN}$} &  \multicolumn{3}{c}{$\mathbf{DBP15K}_{\rm FR\mbox{-}EN}$}    \\
    \cmidrule(r){2-4}\cmidrule(r){5-7}\cmidrule{8-10}
    & H@1  & H@10   &  MRR  
    & H@1  & H@10   &  MRR  
    & H@1  & H@10   &  MRR 
    \\

    \midrule
    
    GCN-align~\cite{GCN-Align}   
    & .434 &  .762  &  .550   
    & .427 &  .762  &  .540  
    & .411 &  .772  &  .530    \\
    
    KECG~\cite{KECG}    
    & .478 &  .835   &  .598   
    & .490 &  .844   &  .610  
    & .486 &  .851   &  .610     \\

    BootEA~\cite{BootEA}  
    & .629 &  .847  &  .703   
    & .622 &  .854   & .701  
    & .653 &  .874   & .731     \\

    NAEA~\cite{NAEA}  
    & .650 &  .867  & .720   
    & .641 &  .873  & .718  
    & .673 &  .894  & .752     \\
    
    \midrule

    EVA~\cite{EVA}   
    & .761 &  .907   &  .814  
    & .762 &  .913   &  .817  
    & .793 &  .942   &  .847     \\
    
    MSNEA~\cite{MSNEA} 
    & .643 &  .865   &  .719  
    & .572 &  .832   &  .660  
    & .584 &  .841   &  .671     \\

    XGEA*~\cite{XGEA}    
    & .803 &  .939   &  .854
    & .794 &  .942   &  .849 
    & .821 &  .954   &  .871     \\    

    UMAEA*~\cite{UMAEA}    
    & .811 &  .969   &  .871
    & .812 &  .973   &  .873 
    & .822 &  .981   &  .884     \\

    MCLEA*~\cite{MCLEA}   
    & .824 &  .955   &  .871
    & .827 &  .959   &  .876 
    & .833 &  .974   &  .886     \\

    DESAlign*~\cite{DESAlign}     
    & .853 &  .967   &  .895   
    & .853 &  .969   &  .898
    & \underline{.866} &  .979   &  .910     \\

    MEAformer*~\cite{Meaformer}     
    & .842 &  .963   &  .887   
    & .842 &  .974   &  .891
    & .829 &  .974   &  .883     \\

    PCMEA*~\cite{PCMEA}   
    & .824 &  .955   &  .872
    & .827 &  .958   &  .876 
    & .847 &  .975   &  .895     \\

    SimDiff~\cite{SimDiff}   
    & .829 &  .963   &  .877
    & .835 &  .966   &  .883 
    & .861 &  .980   &  .905     \\

    IBMEA~\cite{IBMEA}
    & \underline{.859}   &  \underline{.975}   &  \underline{.903}  
    & \underline{.856}    &  \underline{.978}   &  \underline{.902}  
    & .864   &  \underline{.985}   &  \underline{.911}  \\

    \midrule
    \textbf{CDMEA}  
    & \textbf{.865}   &  \textbf{.978}   &  \textbf{.909}  
    & \textbf{.863}    &  \textbf{.983}   &  \textbf{.910}  
    & \textbf{.877}   &  \textbf{.989}   &  \textbf{.922}     \\
    
    \textit{\textcolor{blue}{Improv. \%}}
    & \textcolor{blue}{$\uparrow$0.6}	& \textcolor{blue}{$\uparrow$0.3}	& \textcolor{blue}{$\uparrow$0.6}
    & \textcolor{blue}{$\uparrow$0.7}	& \textcolor{blue}{$\uparrow$0.5}	& \textcolor{blue}{$\uparrow$0.8}
    & \textcolor{blue}{$\uparrow$1.3}	& \textcolor{blue}{$\uparrow$0.4}	& \textcolor{blue}{$\uparrow$1.3}
    \\
    
\bottomrule

\end{tabular}
}
\end{table}

\subsection{Overall Results}\label{sec:main_res}

To verify the effectiveness of CDMEA, we report the overall average results on cross-KG datasets in Table~\ref{fbdb_result} and bilingual datasets in Table~\ref{fbdbp_result}. 
Several key observations are as follows:
(1) \textit{CDMEA consistently outperforms SOTA baselines across 9 benchmarks across three key metrics (H@1, H@10, and MRR).}
Specifically, our model improves H@1 scores for the DBP15K datasets in ${\rm ZH\mbox{-}EN/JA\mbox{-}EN/FR\mbox{-}EN}$ from 0.859/0.856/0.866 to 0.863/0.863/0.877.
Given the high baseline performance, our improvement is still competitive.
Additionally, we observe an average H@1 increase of 2.4\% and 5.1\% on two cross-KG datasets with 80\% and 50\% alignment seed settings, demonstrating model robustness across datasets.
(2) \textit{CDMEA shows stronger performance in low-resource scenarios.}
With only 20\% alignment seeds, our method achieves a 6.80\% increase in H@1 and a 6.85\% improvement in MRR over the runner-up on two cross-KG datasets, indicating the effectiveness of counterfactual debiasing with limited data.
(3) \textit{MMEA methods outperform traditional EA models.}
Our model achieves an average H@1 improvement of 48.4\% (ranging from 38.7\% to 55.8\%) on cross-KG datasets and 21.3\% (ranging from 20.4\% to 22.2\%) on bilingual datasets, highlighting the advantages of incorporating multi-modal information for entity alignment.
Overall, these findings confirm CDMEA’s versatility and reliability across diverse datasets, positioning it as a robust MMEA solution.

\subsection{Ablation Study}\label{sec:ablation_res}

In our ablation study, we evaluate the impact of key components in the CDMEA model on two Cross-KG datasets, as shown in Table~\ref{ablation_result}.
First, removing counterfactual debiasing inference (w/o CDI) reduces H@1 by 3.6\% (FB-DB15K) and 2.4\% (FB-YG15K), and MRR by 2.5\% and 1.5\%, respectively, highlighting CDI’s critical role in mitigating alignment biases, particularly in the visual modality.
However, applying CDI to the graph modality (repl. CDI-G), where biases are minimal, reduces H@1 and MRR by 10.9\% and 8.3\% on average, respectively. 
This suggests that CDI is effective for visual biases but less beneficial in contexts with minimal biases, such as the graph modality.
We also investigate the impact of the loss functions $\mathcal{L}_{v,g,f}$, $\mathcal{L}_v$, $\mathcal{L}_g$, and $\mathcal{L}_f$. 
Excluding $\mathcal{L}_{v,g,f}$ leads to a notable decline in all metrics, underscoring its critical role in guiding the final prediction score toward the optimal target.
The unimodal losses ($\mathcal{L}_v$, $\mathcal{L}_g$, and $\mathcal{L}_f$) marginally reduce performance, confirming the utility of each alignment loss.
Finally, replacing the InfoNCE loss with ICL from prior studies~\cite{MCLEA,Meaformer} leads to moderate performance drops, indicating that InfoNCE better supports model generalization.

\begin{figure}
    \centering
    \includegraphics[width=1\linewidth]{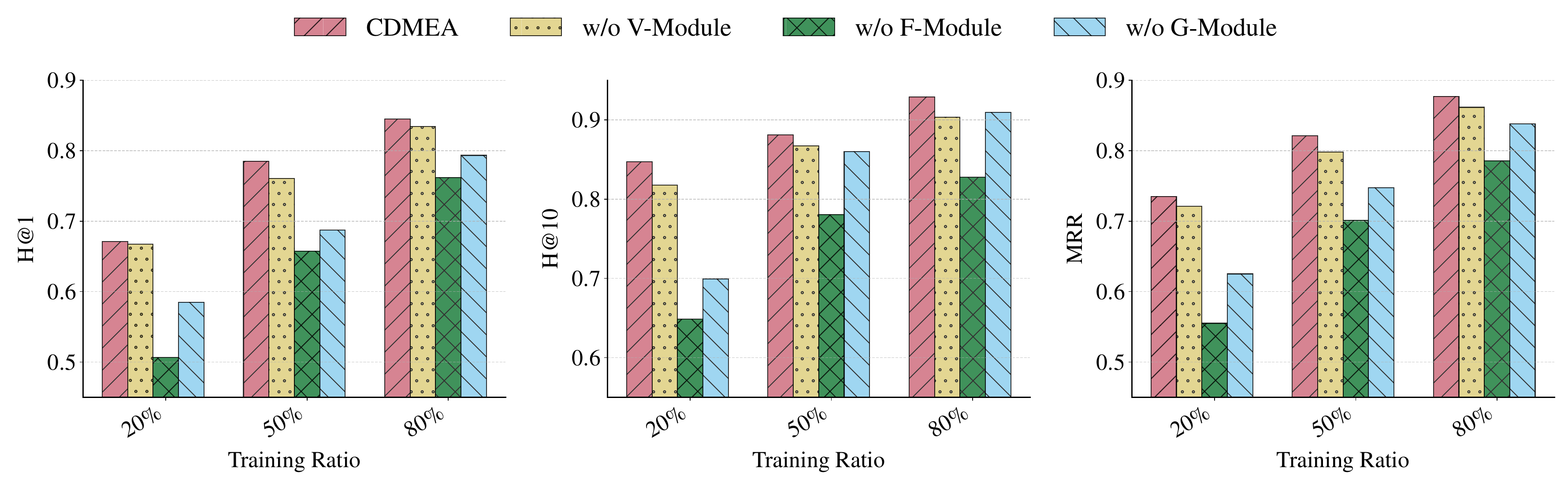}
    \caption{Results of removing the separate modality module from CDMEA on FB-DB15K dataset.}
    \label{fig:diff-modality}
\end{figure}

To evaluate the contribution of each modality branch, we remove the visual modality module (V-Module), fused modality module (F-Module), and graph modality module (G-Module) branches, as shown in Figure~\ref{fig:diff-modality}.
Removing the F-Module results in a substantial performance drop, especially in low-resource settings (20\% training ratio), highlighting the crucial role of the fused modality in exploiting limited data.
Similarly, removing the G-Module causes a large decrease in performance, emphasizing the importance of graph-structured information for MMEA. 
In contrast, removing the V-Module has a relatively minor impact in low-resource settings, as the F-Module effectively uses visual information during fusion. 
Furthermore, in low-resource settings, the V-Module’s influence on MMEA may introduce biased predictions.

\begin{table}[!t]
    \footnotesize
    \centering
    \caption{Variants on two Cross-KG datasets with 20\% alignments seeds. 
    ``\textit{w/o}'' denotes removing the corresponding module from the complete model,
    and ``\textit{repl.}'' means replacing the corresponding module with another module.}
    \label{ablation_result}
        \setlength\tabcolsep{6.0pt}
        {
        \begin{tabular}{@{}rlcccccc@{}}
		
        \toprule

       &  \multirow{2.5}{*}{Model} 
       &  \multicolumn{3}{c}{\textbf{FB-DB15K}}  & \multicolumn{3}{c}{\textbf{FB-YG15K}}  \\ 

        \cmidrule(r){3-5} \cmidrule(r){6-8} 
        &
        & H@1  & H@10   &  MRR  
        & H@1  & H@10   &  MRR 
        \\
        
        \midrule
        \multirow{4}{*}[-5.5ex]{\rotatebox{90}{\textbf{Module}}}
        & CDMEA 
        & \textbf{.674} & \textbf{.861}   & \textbf{.741} 
        & \textbf{.623}   &  \textbf{.811}   &  \textbf{.690} 
        \\
        \midrule

        & \textit{w/o} CDI  
        & .638    & .842  & .716           
        & .599    & .807   & .675  \\

        & \textit{repl.} CDI-G
        & .551    & .797   & .643   
        & .528    & .783   & .622  \\
        
        & \textit{w/o} $\mathcal{L}_{v,g,f}$   
        & .565  & .783  & .646 
        & .537    & .728  & .607   \\

        & \textit{w/o} $\mathcal{L}_{v}$ \& $\mathcal{L}_{g}$ \& $\mathcal{L}_{f}$
        & .644    & .839   & .716   
        & .590    & .800   & .666  \\
        
        & \textit{repl.} ICL
        & .643    & .834   & .713   
        & .581    & .786   & .654  \\
        
        \bottomrule
        \end{tabular}
	}
\end{table}

\begin{table*}[!th]
\footnotesize
\centering
\caption{
Performance comparison of baseline models with counterfactual debiasing inference (CDI) on 2 cross-KG datasets.
}
\label{Baseline_CDI_Fbdb}
\setlength\tabcolsep{4.8pt}
{
{
\begin{tabular}{@{}lccccccccc|ccccccccc@{}}
\toprule
\multirow{2.5}{*}{\bf Methods} & \multicolumn{3}{c}{\bf FB-DB15K (20\%)} & \multicolumn{3}{c}{\bf FB-DB15K (50\%)} &  \multicolumn{3}{c|}{\bf FB-DB15K (80\%)} & \multicolumn{3}{c}{\bf FB-YG15K (20\%)} & \multicolumn{3}{c}{\bf FB-YG15K ((50\%)} &  \multicolumn{3}{c}{\bf FB-YG15K (80\%)}   \\

\cmidrule(r){2-4}\cmidrule(r){5-7}\cmidrule(r){8-10} \cmidrule(r){11-13}\cmidrule(r){14-16}\cmidrule(r){17-19}
& H@1  & H@10   &  MRR  
& H@1  & H@10   &  MRR  
& H@1  & H@10   &  MRR     
& H@1  & H@10   &  MRR     
& H@1  & H@10   &  MRR     
& H@1  & H@10   &  MRR     
\\

\midrule

MCLEA* 
& .447 &  .715   &  .538   
& .626 &  .838   &  .700
& .752 &  .900   &  .807   
& .407 &  .660   &  .494
& .541 &  .762   &  .618 
& .652 &  .847   &  .721     \\

\hspace{1em}\textit{w/} CDI 
& .463	& .719	& .551
& .642	& .843	& .715
& .761	& .915	& .819 
& .432	& .677	& .517
& .551	& .768	& .628
& .668	& .853	& .741 \\

\textit{\textcolor{blue}{Improv. \%}}
& \textcolor{blue}{$\uparrow$1.6}	& \textcolor{blue}{$\uparrow$0.4}	& \textcolor{blue}{$\uparrow$1.3}
& \textcolor{blue}{$\uparrow$1.6}	& \textcolor{blue}{$\uparrow$0.5}	& \textcolor{blue}{$\uparrow$1.5}
& \textcolor{blue}{$\uparrow$0.9}	& \textcolor{blue}{$\uparrow$1.5}	& \textcolor{blue}{$\uparrow$1.2}
& \textcolor{blue}{$\uparrow$2.5}	& \textcolor{blue}{$\uparrow$1.7}	& \textcolor{blue}{$\uparrow$2.3}
& \textcolor{blue}{$\uparrow$1.0}	& \textcolor{blue}{$\uparrow$0.6}	& \textcolor{blue}{$\uparrow$1.0}
& \textcolor{blue}{$\uparrow$1.6}	& \textcolor{blue}{$\uparrow$0.6}	& \textcolor{blue}{$\uparrow$2.0} \\

\cmidrule{1-19}

DESAlign*       
& .541 &  .811   &  .643   
& .706 &  .877   &  .770
& .824 &  .931   &  .863 
& .452 &  .712   &  .547
& .605 &  .802   &  .677 
& .727 &  .876   &  .780  \\

\hspace{1em}\textit{w/} CDI 
& .587	& .816	& .670
& .741	& .881	& .792    
& .827	& .936	& .869    
& .521	& .738	& .599
& .651	& .814	& .709    
& .739	& .880	& .786 \\

\textit{\textcolor{blue}{Improv. \%}}
& \textcolor{blue}{$\uparrow$4.6}	& \textcolor{blue}{$\uparrow$0.5}	& \textcolor{blue}{$\uparrow$2.7}
& \textcolor{blue}{$\uparrow$3.5}	& \textcolor{blue}{$\uparrow$0.4}	& \textcolor{blue}{$\uparrow$2.2}
& \textcolor{blue}{$\uparrow$0.3}	& \textcolor{blue}{$\uparrow$0.5}	& \textcolor{blue}{$\uparrow$0.6} 
& \textcolor{blue}{$\uparrow$6.9}	& \textcolor{blue}{$\uparrow$2.6}	& \textcolor{blue}{$\uparrow$5.2}
& \textcolor{blue}{$\uparrow$4.6}	& \textcolor{blue}{$\uparrow$1.2}	& \textcolor{blue}{$\uparrow$3.2}
& \textcolor{blue}{$\uparrow$1.2}	& \textcolor{blue}{$\uparrow$0.4}	& \textcolor{blue}{$\uparrow$0.6} \\

\cmidrule{1-19}

MEAformer*    
& .558 & .807   &  .647   
& .698 & .871   &  .762  
& .794 & .918   &  .841    
& .449 &  .672   &  .527   
& .593 &  .789   &  .664
& .713 &  .852   &  .766 \\

\hspace{1em}\textit{w/} CDI 
& .637	& .829	& .704
& .758	& .879	& .801    
& .837	& .927	& .869    
& .557	& .733	& .617
& .657	& .801	& .710    
& .737	& .854	& .778 \\

\textit{\textcolor{blue}{Improv. \%}}
& \textcolor{blue}{$\uparrow$7.9}	& \textcolor{blue}{$\uparrow$2.2}	& \textcolor{blue}{$\uparrow$5.7}
& \textcolor{blue}{$\uparrow$6.0}	& \textcolor{blue}{$\uparrow$0.8}	& \textcolor{blue}{$\uparrow$3.9}
& \textcolor{blue}{$\uparrow$4.3}	& \textcolor{blue}{$\uparrow$0.9}	& \textcolor{blue}{$\uparrow$2.8} 
& \textcolor{blue}{$\uparrow$10.8} & \textcolor{blue}{$\uparrow$6.1}	& \textcolor{blue}{$\uparrow$9.0}
& \textcolor{blue}{$\uparrow$6.4}	& \textcolor{blue}{$\uparrow$1.2}	& \textcolor{blue}{$\uparrow$4.6}
& \textcolor{blue}{$\uparrow$2.4}	& \textcolor{blue}{$\uparrow$0.2}	& \textcolor{blue}{$\uparrow$1.2} \\

\bottomrule

\end{tabular}
}}
\end{table*}

\subsection{Further Analysis}

\subsubsection{Generality over Different Baselines}\label{sec:all_baseline}

An important question is whether our model is effective across diverse multi-modal entity alignment baselines.
To this end, we apply our proposed counterfactual debiasing inference (CDI) to three baseline models (MCLEA, DESAlign, MEAformer) on two cross-KG datasets. 
The experimental results are shown in Table~\ref{Baseline_CDI_Fbdb}.
Our proposed counterfactual debiasing inference (w/ CDI) improves H@1 by an average of 1.36\%, 2.80\%, and 6.06\% on FB-DB15K, and 1.70\%, 4.23\%, and 6.53\% on FB-YG15K for MCLEA, DESAlign, and MEAformer, respectively.
MEAformer shows the largest improvement, which can be attributed to its cross-modal fusion structure during training, aligning closely with CDI's debiasing mechanism during inference.
Experimental results also indicate that the inclusion of CDI leads to greater performance improvements under low-resource scenarios. 
We believe that CDI can mitigate the model bias issue introduced by limited training data, thereby enhancing entity alignment performance.

\subsubsection{Robustness under Low-resource Scenarios}\label{sec:low_res}

\begin{figure}
    \centering
    \includegraphics[width=1\linewidth]{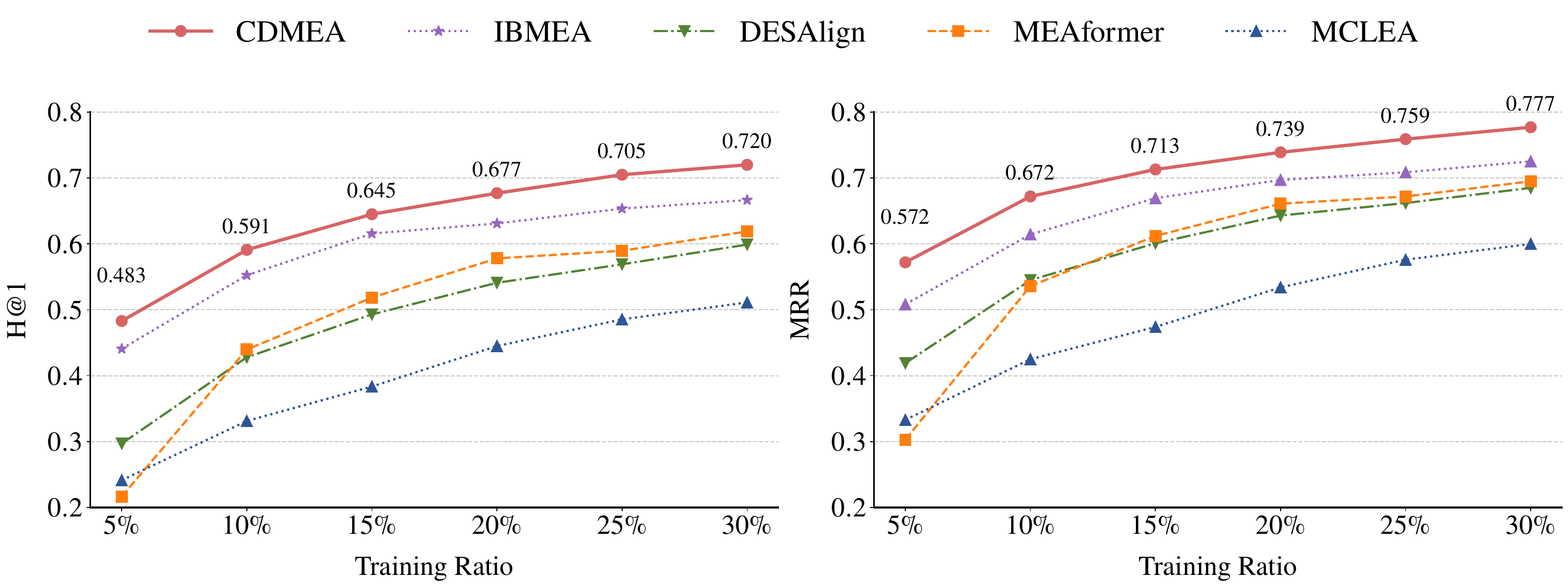}
    \caption{Results in the low-resource data scenario with different training ratios of seed alignments on FB-DB15K dataset.}
    \label{fig:low-ratio}
\end{figure}

Since the model may overfit on misleading cues with few training data and exacerbate modality bias, we conduct experiments under low-resource scenarios with training data ratios ranging from 5\% to 30\%. 
We evaluate IBMEA, MEAformer, DESAlign, and MCLEA as baselines, with results shown in Fig.~\ref{fig:low-ratio}.
As alignment seed ratios decrease, all models experience varying levels of performance degradation, primarily due to insufficient training data, which increases the likelihood of biased predictions during inference.
Notably, CDMEA is less affected by reduced training data.
With 5\% of the training data, CDMEA performs similarly to DESAlign and MEAformer at 10\%, showing its robustness.
This is attributed to CDMEA’s CDI, which effectively mitigates visual modality bias in low-resource settings.

\subsubsection{Robustness under High-noise Image Scenarios}\label{sec:high-noise}
\begin{figure}
    \centering
    \includegraphics[width=0.99\linewidth]{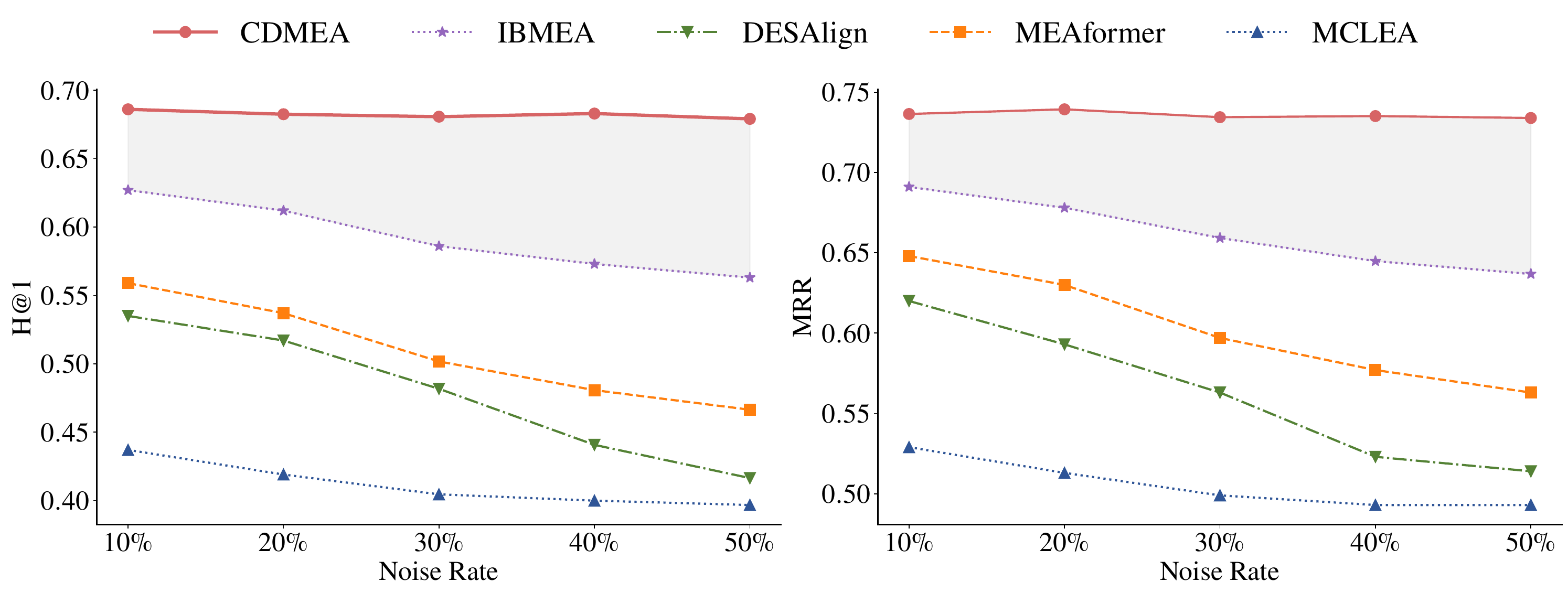}
    \caption{Results on noisy image samples in the FB15K-DB15K dataset with random dropout at varying rates.}
    \label{fig:high-noise}
\end{figure}

To evaluate the robustness of CDMEA under noisy conditions, we use the FB-DB15K dataset (with 20\% seed alignments) and introduce artificial noise to the images via random dropout at varying rates. 
As shown in Figure~\ref{fig:high-noise}, all baseline models exhibit a performance decline as the image noise rate increases, while our model maintains relatively high H@1 and MRR scores.
Considering that higher noise rates tend to generate more modal bias, the results demonstrate that our causal debiasing model can effectively mitigate the introduced modal bias and maintain very high robustness in noisy scenarios.

\subsubsection{Robustness under Low-similarity Image Scenarios.}
To explore the robustness of CDMEA with hard images, we consider entity pairs with low image similarity, discussed in Section~\ref{Sec:intro}.
We train the model using 20\% of the training data from the FB-DB15K dataset and select entities with varying levels of image similarity from the test set.
As shown in Figure~\ref{fig:intro_low_img}, our model achieves the best performance on both the H@1 and MRR metrics across different image similarity settings.
Notably, the results improve as image similarity decreases, suggesting that CDMEA is effective in mitigating modality bias in low-similarity image scenarios.

\begin{figure}
    \centering
    \includegraphics[width=1\linewidth]{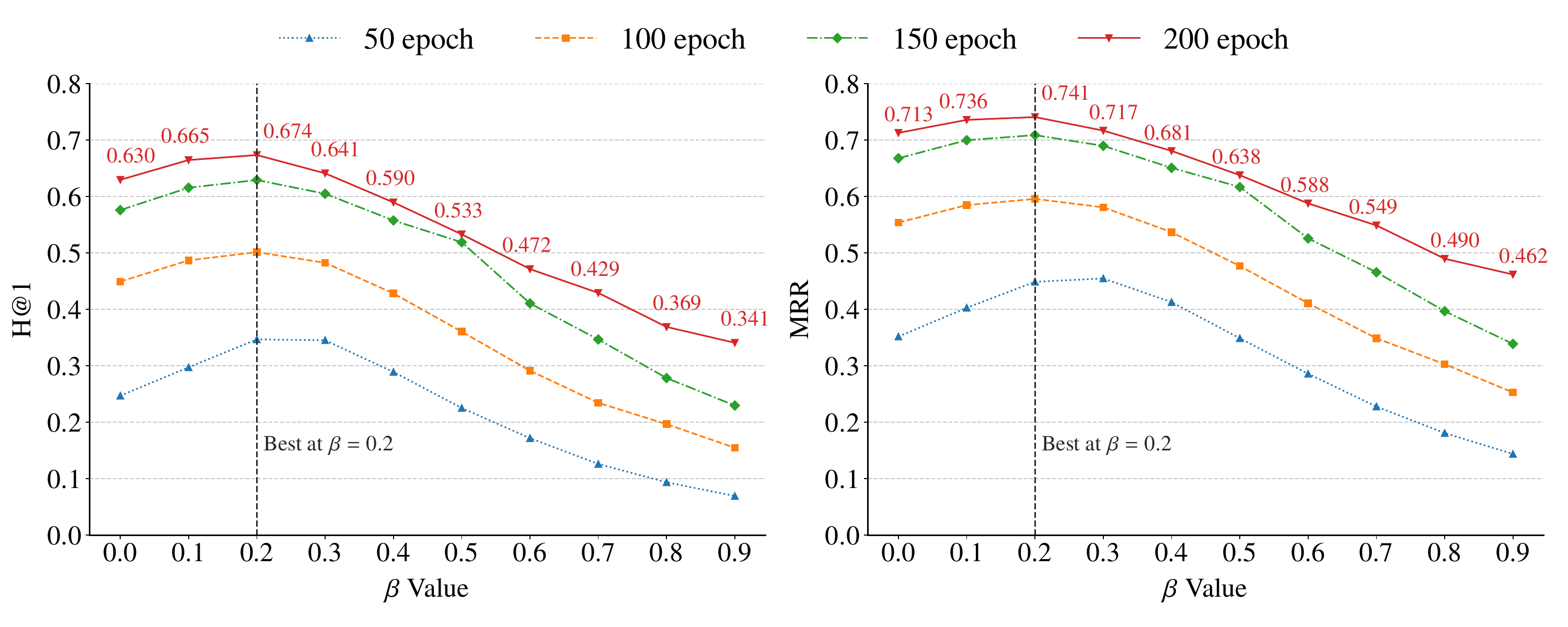}
    \caption{Results of TIE with varying $\beta$ at different epochs.}
    \label{fig:diff-alpha}
\end{figure}

\subsubsection{Impact of the $\beta$ Parameter on TIE}\label{sec:alpha}

To further evaluate the impact of the hyperparameter $\beta$ in Eq.~(\ref{eq:debias_inference}), we conduct experiments on the FB-DB15K dataset with 20\% seed alignments and varying training epochs. 
The hyperparameter $\beta$ controls the trade-off between the total indirect effect (TIE) and the total effect (TE). 
At $\beta = 0$, TIE equals TE, corresponding to traditional inference based on posterior probabilities. 
As $\beta$ increases from 0 to 0.2, performance improves, peaking at $\beta = 0.2$, and then declines beyond this point (Fig.~\ref{fig:diff-alpha}). 
These results suggest that balancing visual information with bias minimization is optimal for MMEA.
Notably, at $\beta = 0.2$, the model significantly improves H@1 and MRR, especially during early training stages (epochs $\le$ 150). 
This demonstrates that debiased inference applied to visual information improves prediction accuracy and accelerates model convergence during training.

\subsubsection{Efficiency Analysis}\label{sec:effciency}
To evaluate the efficiency of our CDMEA model, we compare its training time and parameter size against four baselines (MCLEA, MEAformer, DESAlign, IBMEA) on the FB-DB15K dataset with 20\% seed alignments.
The parameter sizes are 8.9M, 10.5M, 11.4M, 10.9M, and 11.1M for the baselines and CDMEA, respectively.
Training times are 6,018s, 5,764s, 9,540s, 1,797s, and 1,393s.
CDMEA achieves a 22.4\% reduction in training time compared to IBMEA and speedups of  4.32x, 4.13x, and 6.84x over MCLEA, MEAformer, and DESAlign.
This improvement is due to CDMEA's CDI mechanism, which mitigates visual modality bias for hard samples, allowing faster convergence and fewer iterations, offering a clear advantage for low-computation scenarios.

\subsection{Qualitative Analysis}\label{sec:case}

\begin{figure}
    \centering
    \includegraphics[width=0.85\linewidth]{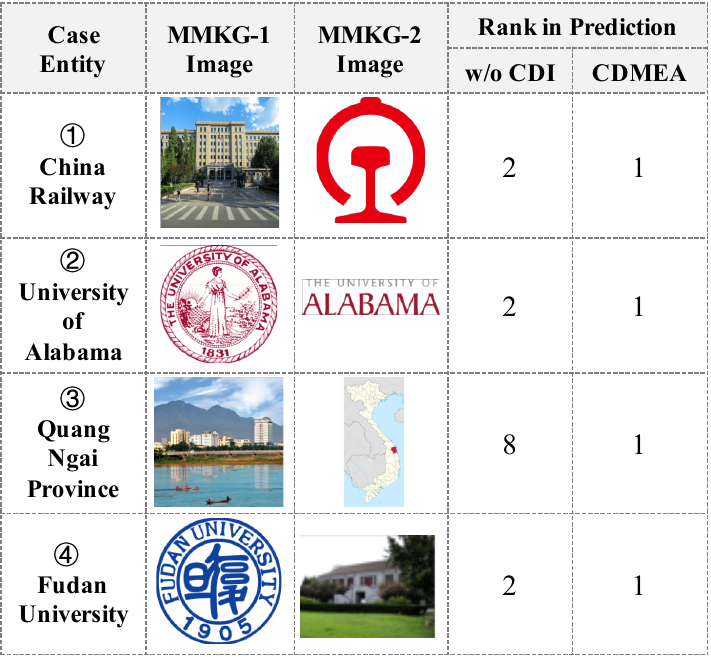}
    \caption{Examples of equivalent entities with images of different styles. w/o CDI denotes the model without counterfactual debiasing inference.}
    \label{fig:CDMEA-case}
\end{figure}

To demonstrate CDMEA’s ability to mitigate visual bias in MMEA, we present examples from $\mathrm{DBP15K}_{\rm ZH\mbox{-}EN}$ in Fig.~\ref{fig:CDMEA-case}, each showing entities with images from two MMKGs (MMKG-1 and MMKG-2).
In Case 1, \textit{China Railway} with an office building in MMKG-1 and a logo in MMKG-2. 
Case 2 features the \textit{University of Alabama}, with its emblem in MMKG-1 and a modern logo in MMKG-2. 
Case 3 shows \textit{Quang Ngai Province} is depicted with a landscape in MMKG-1 and a map in MMKG-2. 
Case 4 highlights \textit{Fudan University}, with its emblem in MMKG-1 and campus buildings in MMKG-2.
Without counterfactual debiasing inference (w/o CDI), the model misaligns entities, leading to poor ranks (e.g., rank 8 for \textit{Quang Ngai Province}, rank 2 for others). 
In contrast, CDMEA consistently improves alignment, achieving rank 1 in all cases, especially when the same entity has images from different contexts (e.g., \textit{Quang Ngai Province}'s map and landscape, or mismatched logos and emblems for other entities).
These results show that CDMEA mitigates visual biases and enhances entity alignment by leveraging multi-modal information.

\section{Related Work}\label{sec:related_works}

\subsection{Entity Alignment}
Entity Alignment (EA) is a typical IR problem~\cite{SEA_sigir,EasyEA_sigir,ER_sigir}, which aims to retrieve equivalent entities between KGs to facilitate knowledge fusion and entity retrieval. 
Most EA methods focus on traditional KGs, which are represented by relational triples and graph structures.
These methods can be categorized into two groups:
(1) \textit{Structure-based approaches} including knowledge embedding methods ~\cite{MTransE,Transedge,IPTransE,BootEA} and graph-based models ~\cite{GCN-Align,KECG,RDGCN,MuGNN,AliNet,SEA_sigir},  focus on capturing structural information through relational triples or by aggregating neighborhood features.
(2) \textit{Side information-based approaches} incorporate additional side information (e.g., entity names, attributes, and relation predicates) to enhance entity representations~\cite{JAPE,AttrE,MultiKE,KDCoE,AttrGNN,EasyEA_sigir}.
While structure-based methods assume aligned entities share similar neighborhoods, and side information-based methods rely on auxiliary data, both struggle to integrate visual-modality data, limiting performance in MMEA.

\subsection{Multi-modal Entity Alignment}
With the rise of MMKGs, Multi-modal Entity Alignment (MMEA) has gained significant attention for multi-modal knowledge-based IR applications.
MMEA improves EA accuracy by integrating visual and other modalities.
Early approaches~\cite{POE,MMEA,HMEA,MSNEA} utilize direct or fixed-weight operations to combine multi-modal information but lack adaptability in learning the relative importance of each modality.
To address this, attention-based methods~\cite{EVA,MCLEA,ACK-MMEA} incorporate learnable weighted attention to model modality importance.
Graph-based methods~\cite{LoginMEA,XGEA} use GNNs to combine cross-modal information, improving entity relation understanding.
Further performance boosts come from dynamic pseudo-label generation in PSNEA~\cite{PSNEA} and PCMEA~\cite{PCMEA}, while SimDiff~\cite{SimDiff} uses diffusion-based data augmentation~\cite{Diffusion_severy} to stabilize training in low-resource settings.
Contrastive learning is employed in MCLEA~\cite{MCLEA}, and UMAEA~\cite{UMAEA} addresses missing modalities. 
Additionally, MEAformer~\cite{Meaformer} extends UMAEA by dynamically learning modality weights through transformers, DESAlign~\cite{DESAlign} reduces semantic inconsistency with Dirichlet energy, and IBMEA~\cite{IBMEA} applies the information bottleneck principle to handle misleading cues.
Despite the advancements made by these methods, they often overlook the potential impact of visual modality bias on the final MMEA predictions.

\section{Conclusion}
This paper investigates the visual modality bias in MMEA from a causal perspective and proposes a general counterfactual debiasing framework for MMEA, termed CDMEA.
We abstract out a causal graph of MMEA, and achieve the CDMEA with advanced implementations.
In this way, we take advantage of both visual and graph modalities, and suppress the direct causal effect of the visual modality, preventing the bias towards entity images.
Experimental results indicate that CDMEA outperforms 14 previous state-of-the-art baselines, and shows remarkable results in the low-similarity, high-noise, and low-resource data scenarios.

\begin{acks}
This work is supported by the National Natural Science Foundation of China (No.62406319), the Youth Innovation Promotion Association of CAS (No.2021153), and the Postdoctoral Fellowship Program of CPSF (No.GZC20232968).
\end{acks}

\bibliographystyle{ACM-Reference-Format}
\bibliography{fp1215}










\end{document}